\documentclass[aps, prl, twocolumn, amsmath,amssymb]{revtex4}

\usepackage{tikz}
\usepackage{graphicx}
\usepackage{dcolumn}
\usepackage{bm}

\newtheorem{theorem}{Theorem}
\newtheorem{definition}[theorem]{Definition}

\newtheorem{lemma}[theorem]{Lemma}

\def\Tr{\textnormal{Tr}}
\def\Mat{\textnormal{Mat}}
\def\Re{\textnormal{Re}}

\def\ls{\hskip0.03in}

\begin{document}

\title{Impossibility of perfect cheating for single-qubit position verification}

\author{Carl A.~Miller}
 \affiliation{National Institute of Standards and Technology, Gaithersburg, MD, USA}
 \affiliation{Joint Center for Quantum Information and Computer Science (QuICS), University of Maryland, College Park, MD, USA}
\author{Yusuf Alnawakhtha}
 \affiliation{Department of Computing, Arabian Gulf University, Manama, Bahrain}
 \affiliation{Joint Center for Quantum Information and Computer Science (QuICS), University of Maryland, College Park, MD, USA}

\date{\today}

\begin{abstract}
In quantum position verification (QPV), a prover certifies her location to a set of verifiers by performing a quantum computation. One of the first QPV protocols was proposed by Kent, Munro, and Spiller in 2011: the prover receives a qubit $Q$ from one direction, receives an orthogonal basis $\{ v, v^\perp \}$ from the opposite direction, then measures $Q$ in $\{ v, v^\perp \}$ and broadcasts the result.  Several variants of this protocol have been proposed and analyzed, but the question of whether the original protocol is secure has never been fully resolved. In this work we show that there is no perfect finite-dimensional cheating strategy for the single-qubit measurement protocol.  Our proof uses tools from real algebraic geometry.
\end{abstract}

\maketitle

Digital communication often requires one party to authenticate another at a distance.  While secret keys and biometrics are credentials commonly used for authentication, one can imagine situations in which a party's spatial position is also a natural credential for participating in a protocol. Consider, for example, communication with a diplomat residing at an embassy: one might want to confirm that the messages received have been issued from the geographical location of the embassy as an added layer of security. 
Constructing secure protocols for position verification is challenging:  there is an efficient known attack~\cite{chandran2009position} which can be applied to any position verification protocol that is based on exchanging classical information. This attack requires the adversaries to make copies of the messages they receive from the verifiers. Quantum Position Verification (QPV) \cite{kent2006patent}
utilizes the no-cloning principle to bar such attacks, and offers the potential for a higher level of security in position-based authentication.

Among the protocols proposed in the early papers on QPV \cite{kent2011quantum, Buhrman_2014, Beigi_2011, malaney2010location, Tomamichel_2013, unruh2014quantum}, one of the simplest --- and arguably the least demanding in terms of quantum resources --- is a protocol from \cite{kent2011quantum} which we will refer to as the \textbf{single-qubit measurement protocol}.  Suppose that a prover $\mathbf{P}$ is located between two verifiers $\mathbf{V}_1$ and $\mathbf{V}_2$.  The verifiers secretly agree on a random one-dimensional projector $P$ on $\mathbb{C}^2$.  The first verifier $\mathbf{V}_1$ transmits $P$ to $\mathbf{P}$, while the second verifier $\mathbf{V}_2$ prepares a qubit by encoding a bit $z$ into the eigenbasis for $P$ and simultaneously transmits the qubit to $\mathbf{P}$. The prover $\mathbf{P}$ recovers $z$ by applying the measurement $\{ P, \mathbb{I} - P \}$ to the qubit (where $\mathbb{I}$ denotes the identity matrix) and transmits $z$ back to $\mathbf{V}_1$ and $\mathbf{V}_2$.  The verifiers goal is to prove that $\mathbf{P}$'s responses in this protocol could not be faked by any parties who are not at $\mathbf{P}$'s purported location, even if multiple adversaries were to cooperate from different points in space.

In more than a decade of activity on the theory of QPV, various additional QPV protocols have been proposed and cheating strategies have been studied in detail~\cite{qi2015loss,  liu2021beating,bluhm2022single,junge2022geometry,allerstorfer2023security,allerstorfer2024continuous}. However, the full security of the simple single-qubit measurement scheme has remained an open question. Lau and Lo \cite{lau2011insecurity} showed that if the possible choices for $P$ are restricted to $\left\{ \left| 0 \right> \left< 0 \right|, \left| + \right> \left< + \right| \right\}$ (the ``BB84'' case) then two adversaries can cheat by sharing a single maximally-entangled qubit pair.  They then considered the more general case in which $P$ can be arbitrary, and they showed that, in contrast, cheating strategies based on shared entangled qubits  or qutrits cannot exist.  
Chakraborty and Leverrier \cite{chakraborty2015practical} showed that if $P$ is restricted to the $k$th level of the Clifford hierarchy then a perfect finite-dimensional cheating strategy exists, although it requires an amount of entanglement that grows with $k$.  More recently, Olivo et al.~\cite{olivo2020breaking} found new cheating strategies for basis sets of the form $\{ P_0, P_\theta \}$, where $P_\theta$ denotes the projector onto $(\cos \theta) \left| 0 \right> + (\sin \theta ) \left| 1 \right>$.

Other attacks have been designed to cheat at certain families of QPV protocols~\cite{buhrman2013garden, speelman2015instantaneous, allerstorfer2022role, dolev2022non,cree2023code} by exploiting particular structures of the protocols. General QPV cheating results, such as those based on port-based teleportation \cite{Beigi_2011}, show that adversaries can \textit{approximately} cheat at \emph{any} QPV protocol, up to an arbitrarily small error term, by pre-sharing an amount of entanglement that increases as the error term shrinks.  But, these results leave a central question unanswered: does there exist a general perfect cheating strategy for the single-qubit measurement protocol?  In the current work, we answer this question in the negative. We prove that any cheating strategy that is based on a finite-dimensional entangled system can only succeed perfectly for a finite number of possible basis choices (see Theorem~\ref{thm:main}).  
Our result shows that requiring the prover to measure the received qubit in an arbitrary basis -- even in the absence of any other techniques designed to guard against cheating -- is enough to rule out the possibility of a perfect attack by finite-dimensional adversaries.
Our proof uses a tool from real algebraic geometry (the Milnor-Thom theorem), providing a new approach for proving QPV security results.
This suggests that incorporating algebraic geometry techniques could open up new directions for developing QPV protocols with stronger security guarantees.

\section{Definitions and Notation}

\label{sec:def}

A \textbf{quantum register} is a complex Hilbert space $\mathcal{H}$ together with a fixed isomorphism $\mathcal{H} \cong \mathbb{C}^n$, with $n \geq 1$.  The image of the standard basis for $\mathbb{C}^n$ under this isomorphism is referred to as the standard basis for $\mathcal{H}$. The expression $\mathbb{I}_\mathcal{H}$ denotes the identity map on $\mathcal{H}$. If $\mathcal{H}, \mathcal{J}$ are quantum registers, then we may write $\mathcal{H}  \mathcal{J}$ for the tensor product $\mathcal{H} \otimes \mathcal{J}$.  
If  $c$ is an element of $\mathcal{H} \otimes \mathcal{J}$ and $c = \sum_{ij} c_{ij} e_i \otimes f_j$, where $e_i$ and $f_j$ denote standard basis vectors from $\mathcal{H}$ and $\mathcal{J}$ respectively and $c_{ij} \in \mathbb{C}$, then  
$\Mat(c)$ denotes the matrix 
\begin{equation}
\Mat(c)  =  \left[ c_{ij} \right]_{ij}.
\end{equation}
Note that $\Mat(c)$ determines a linear map from $\mathcal{J}$ to $\mathcal{H}$.
The expression $L ( \mathcal{H} )$ denotes the set of all linear maps from $\mathcal{H}$ to itself. For any $u, v \in \mathcal{H} \otimes \mathcal{J}$, the following two identities hold (Proposition $1$ of~\cite{Gutoski2007toward}):
\begin{eqnarray}
    \Tr_{\mathcal{J}}(u v^*) = \Mat(u) \Mat(v)^* \label{eq: partial trace of second system}
\end{eqnarray}
\begin{eqnarray}    
    \Tr_{\mathcal{H}}(u v^*) = \overline{\Mat(u)^* \Mat(v)}, \label{eq: partial trace of first system}
\end{eqnarray}
where $M^*$ denotes the adjoint (i.e., the conjugate transpose) of a linear map $M$.

An \textbf{analog register} is a subset $D \subseteq \mathbb{R}^n$ of a finite-dimensional real vector space.  We refer to the elements of $D$ as the \textbf{states} of the analog register.

A \textbf{qubit} is a quantum register of dimension $2$.  Let $\mathbb{S}$ denote the set of all one-dimensional orthogonal projectors on $\mathbb{C}^2$ (the Bloch sphere).  Every element of $\mathbb{S}$ can be uniquely expressed in terms of Pauli operators as $\left(  \mathbb{I} + a X + bY + cZ \right)/2$,
where $\mathbb{I}$ denotes the $2\times 2$ identity matrix, $(a,b,c) \in \mathbb{R}^3$ is a unit vector, and 
\begin{eqnarray}
    \begin{array}{ccc}
X = \left[ \begin{array}{cc} 0 & 1 \\
1 & 0 \end{array} \right], &
Y = \left[ \begin{array}{cc} 0 & i \\
-i & 0 \end{array} \right], &
Z = \left[ \begin{array}{cc} 1 & 0 \\
0 & -1 \end{array} \right].
    \end{array} 
\end{eqnarray}

A \textbf{pure state} of a quantum register $\mathcal{A}$ is a unit vector in $\mathcal{A}$. A \textbf{density operator} on $\mathcal{A}$ is positive semidefinite linear operator of trace $1$.
A \textbf{quantum channel} $\Psi$ from a quantum register $\mathcal{V}$ to a quantum register $\mathcal{W}$ is a completely positive trace-preserving map from $L ( \mathcal{V})$ to $L ( \mathcal{W})$.  Such a channel is an \textbf{isometric} channel if it is of the form $\Psi ( M ) = U M U^*$ where $U \colon \mathcal{V} \to \mathcal{W}$ is an isometry.

Two density matrices $\rho, \chi$ on a quantum register $\mathcal{A}$ are \textbf{perfectly distinguishable} if the support of $\rho$ is orthogonal to the support of $\chi$.  Two density matrices  $\alpha, \beta$ on a bipartite quantum register $\mathcal{A} \otimes \mathcal{B}$ are \textbf{perfectly distinguishable on $\mathcal{B}$} if the support of $\Tr_\mathcal{A} \alpha$ in $\mathcal{B}$ is orthogonal to that of $\Tr_\mathcal{A} \beta$. 

A subset $V \subseteq \mathbb{R}^m$ is \textbf{open} if for any $v \in V$, there exists a ball of radius $\epsilon > 0$ centered on $v$ that is also contained in $V$.  A subset of $\mathbb{R}^m$ is \textbf{closed} if its complement is open.  A closed subset $W \subseteq \mathbb{R}^m$ is \textbf{connected} if it cannot be expressed as the disjoint union of two nonempty closed sets.  A \textbf{connected component} of a closed set $Z \subseteq \mathbb{R}^m$ is a nonempty subset of $Z$ that is closed and connected.  Two subsets $U \subseteq \mathbb{R}^n$ and $V \subseteq \mathbb{R}^m$ are \textbf{diffeomorphic} if there exists a bijective map $F \colon U \to V$ such that $F$ and $F^{-1}$ are infinitely differentiable.

We state a form of the Milnor-Thom theorem that is central to the proof of our main result.
The following theorem is a direct corollary of Theorem 2 in~\cite{milnor1964betti}.
\begin{theorem}
    Let $n, k \in \mathbb{N}$ and let $\Lambda$ be a subset of $\mathbb{R}^n$ that is defined by polynomial equations of degree at most $k$.
    The number of connected components of $\Lambda$ is at most $k(2k-1)^{n-1}$.
\end{theorem}

\section{The Single-Qubit Measurement Protocol}

\label{sec:kms}

The single-qubit measurement protocol is a one-dimensional position verification protocol, so we assume that all parties involved are located on a single line with a distinguished point (the origin).  The line is parameterized by a real value $x \in \mathbb{R}$, so that $x = \ell$ refers to the point $\ell$ meters to the right of the origin (or $-\ell$ meters to the left of the origin, if $\ell$ is negative).  Time is measured in seconds.  (We will generally omit units.)

Let the two verifiers $\mathbf{V}_1$ and $\mathbf{V}_2$ be located at $x = -d$ and $x = d$, respectively, where $d$ is a positive real parameter.  
An honest prover $\mathbf{P}$ is located at $x = 0$ \footnote{For simplicity, we assumed that the verifiers are checking that a prover is exactly at the midpoint between them. In general, they can check that a prover is at any point between them by having one of the verifiers delay their message in Step~\ref{verifiers transmission} such that the verifiers' messages arrive at the target position at the same time.}.
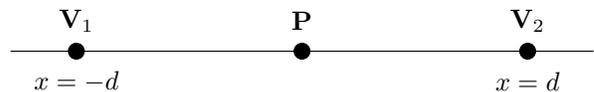
\begin{eqnarray*}
\begin{tikzpicture}
    \node (start) at (-1,0) {};
    \node (finish) at (7,0) {};
    \draw (start) -- (finish) {};
    \filldraw (0,0) circle (3pt) {};
    \filldraw (3,0) circle (3pt) {};
    \filldraw (6,0) circle (3pt) {};
    \node (vone) at (0,0.4) {$\mathbf{V}_1$};
    \node (p) at (3,0.4) {$\mathbf{P}$};
    \node (vtwo) at (6,0.4) {$\mathbf{V}_2$};
    \node (label1) at (0,-0.4) {$x=-d$};
    \node (label2) at (6,-0.4) {$x=d$};    
\end{tikzpicture}
\end{eqnarray*}
The protocol proceeds as follows.  Let $c$ denote the speed of light.  
\begin{enumerate}
\item The verifiers choose an element $P \in \mathbb{S}$ and a bit $z \in \{ 0, 1 \}$.

\item At time $t = 0$, the verifier $\mathbf{V}_1$ transmits $P$ to the right (in an analog register).  The verifier $\mathbf{V}_2$ prepares a qubit $\mathcal{Q}$ in state $P$ if $z = 0$, and in state $\mathbb{I} - P$ if $z = 1$, and transmits $\mathcal{Q}$ to the left. \label{verifiers transmission}

\item At time $t = d/c$, the prover $\mathbf{P}$ applies the measurement $\left\{ P , \mathbb{I} - P \right\}$ to the qubit $\mathcal{Q}$, and broadcasts the result.

\item At time $t = 2d/c$, the verifiers $\mathbf{V}_1$ and
$\mathbf{V}_2$ receive bit messages (from $\mathbf{P}$) which we denote by $z_1$ and $z_2$, respectively.  The verifiers check whether the bits $z, z_1, z_2$ all agree; if they do, the verifiers ACCEPT.  Otherwise, they ABORT.
\end{enumerate}

Next, we mathematically characterize cheating strategies for the single-qubit measurement protocol. As is standard in QPV, we assume a cheating model in which there are exactly two adversaries, one on each side of $x=0$.  
We label the adversaries as $\mathbf{A}$ (Alice), located at $x = -h$, and $\mathbf{B}$ (Bob), located at $x = h$, where $h$ is  a positive real number less than $d$.
(Cheating strategies where $h > d$ or that involve more than $2$ adversaries can be simulated by strategies of this form.)
\begin{eqnarray*}
\begin{tikzpicture}
    \node (start) at (-1,0) {};
    \node (finish) at (7,0) {};
    \draw (start) -- (finish) {};
    \filldraw (0,0) circle (3pt) {};
    \filldraw (2,0) circle (3pt) {};
    \filldraw (4,0) circle (3pt) {};
    \filldraw (6,0) circle (3pt) {};
    \node (vone) at (0,0.4) {$\mathbf{V}_1$};
    \node (a) at (2,0.4) {$\mathbf{A}$};
    \node (b) at (4,0.4) {$\mathbf{B}$};
    \node (vtwo) at (6,0.4) {$\mathbf{V}_2$};
\end{tikzpicture}
\end{eqnarray*}

\begin{definition}
\label{def:cheat}
A \textbf{cheating strategy} for a subset $T \subseteq \mathbb{S}$ consists of the following data.
\begin{enumerate}
    \item Quantum registers $\mathcal{A}, \mathcal{B}, \mathcal{C}, \mathcal{D}$.

    \item A pure quantum state $\psi \in \mathcal{A} \otimes \mathcal{B}$.

    \item For every $P \in T$, an isometry $U_P \colon \mathcal{A} \to \mathcal{A} \otimes \mathcal{C}$.  

    \item An isometry $V \colon \mathcal{B} \otimes \mathcal{Q} \to \mathcal{B} \otimes \mathcal{D}$.   
\end{enumerate}
\end{definition}
The systems $\mathcal{A}$ and $\mathcal{B}$ represent Alice and Bob's local memories, respectively.  It is assumed that before the protocol begins, Alice and Bob prepare $\mathcal{A}$ and $\mathcal{B}$ in the state $\psi$, and the verifier $\mathbf{V}_2$ chooses a random bit $z$ and encodes it into $\mathcal{Q}$ in the basis $\{ P , \mathbb{I} - P \}$.
At time ${(d-h)/c}$, Alice receives $P$, applies the isometry $U_P$, and sends the system $\mathcal{C}$ to Bob.  She also re-broadcasts $P$.  Also at time $(d-h)/c$, Bob receives $\mathcal{Q}$, applies the isometry $V$ to $\mathcal{B}$ and $\mathcal{Q}$, and sends the system $\mathcal{D}$ to Alice. 

At time $(d+h)/c$, Alice is in possession of the systems $\mathcal{A}$ and $\mathcal{D}$, and Bob is in possession of $\mathcal{B}$ and $\mathcal{C}$.  We make the following definition.

\begin{definition}\label{def: perfect cheating}
    A cheating strategy for a subset $T \subseteq \mathbb{S}$ (as in Definition~\ref{def:cheat}) is \textbf{perfect} if for every $P \in T$, the states
    \begin{eqnarray}
    \label{perfdist}
        V U_P (\psi \otimes P) \textnormal{ and }
        V U_P ( \psi \otimes (\mathbb{I} - P ))
    \end{eqnarray}
    are perfectly distinguishable on $\mathcal{A} \mathcal{D}$ and perfectly distinguishable on $\mathcal{B} \mathcal{C}$.
\end{definition}
In other words, the cheating strategy is perfect if at time $(d+h)/c$ there are local measurements that Alice and Bob can perform that will allow them to both perfectly guess $z$.

\section{Hidden Measurement Channels}

\label{sec:hid}

This section analyzes the channel that the adversaries would need to apply before they communicate their messages to each other. Note that the single-qubit measurement protocol consists of the verifiers encoding a bit $b$ in the basis $\{ P, \mathbb{I} - P \}$, then providing the prover with the encoded bit and the basis of the encoding. 
The prover is then expected to respond with the bit $b$ to both verifiers. Hence, for the adversaries to cheat at the protocol, they must each be able to recover $b$ from their local registers. This motivates the following channel definition. 

\begin{definition}
\label{def:hid}
Let $\mathcal{Q}$ be a qubit register, and let $P$ be an element of the Bloch sphere $\mathbb{S}$.  A quantum channel
\begin{eqnarray*}
    \Phi \colon L ( \mathcal{Q} )
    \to L ( \mathcal{V}_1 \otimes \mathcal{V}_2 ),
\end{eqnarray*}
(where $\mathcal{V}_1, \mathcal{V}_2$ are quantum registers) is a \textbf{hidden measurement channel} for $P$ if the states $\Phi (P )$ and  $\Phi ( \mathbb{I} - P )$ are perfectly distinguishable on $\mathcal{V}_1$ and perfectly distinguishable on $\mathcal{V}_2$.
\end{definition}
Such channels allow $b$ to be recovered from both systems $\mathcal{V}_1$ and $\mathcal{V}_2$. However, it may not be obvious \textit{how} to recover $b$ from either system. This is the reason for calling such a channel a ``hidden'' measurement channel.

The following theorem considers channels like those in Definition~\ref{def:hid} but with an additional input register representing the adversaries' entangled systems.

\begin{theorem}
\label{thm:finitevalues}
Let
\begin{eqnarray*}
    \Psi \colon L ( \mathcal{W} \otimes \mathcal{Q} ) \to L ( \mathcal{V}_1 \otimes \mathcal{V}_2 )
\end{eqnarray*}
be a quantum channel, where $\mathcal{W}, \mathcal{V}_1, \mathcal{V}_2$ are quantum registers, and $\mathcal{Q}$ is a qubit register.  
Let $\Lambda$ be the set of all pairs $(P, w)$ where $P \in \mathbb{S}$ and $w$ is a unit vector in $\mathcal{W}$, such that the quantum channel from $\mathcal{Q}$ to $\mathcal{V}_1 \otimes \mathcal{V}_2$ defined by 
\begin{eqnarray*}
    \rho \mapsto \Psi ( ww^* \otimes \rho )
\end{eqnarray*}
is a hidden measurement channel for $P$. Let $n = \dim ( \mathcal{W})$. Then, at most $4 \cdot 7^{2n+2}$ distinct values of $P \in \mathbb{S}$ occur in the pairs contained in $\Lambda$.
\end{theorem}

\textit{Proof.} To prove this theorem, we begin by describing polynomial constraints which all elements of $\Lambda$ must satisfy.
Then, in Lemma~\ref{innerlemma}, we prove that for any two pairs $(P, v)$ and $(L,w)$ in $\Lambda$, the distance between $P$ and $L$ is upper bounded by a quadratic function of the distance between $v$ and $w$.
We use this to show, in Lemma~\ref{lem:diffpath}, that the $\mathbb{S}$-values must be the same for all pairs in any connected component of $\Lambda$.
Finally, we use the Milnor-Thom theorem to upper bound the number of connected components and hence the number of distinct values of $P \in \mathbb{S}$ that occur in elements of $\Lambda$.

It suffices to prove the theorem under the assumption that $\Psi$ is an isometric channel $\Psi ( \cdot) = U ( \cdot ) U^*$ defined by a linear map
\begin{eqnarray}
U \colon \mathcal{W} \otimes \mathcal{Q} \to \mathcal{V}_1 \otimes \mathcal{V}_2,
\end{eqnarray}
satisfying $U^*U = \mathbb{I}$, so we make that assumption for the remainder of the proof \footnote{The Stinespring dilation theorem (see section 2.2 in \cite{watrous2018theory}) implies that any channel $\Psi$ from $\mathcal{WQ}$ to $\mathcal{V}_1\mathcal{V}_2$ can be expanded to an isometric channel $\Psi'$ from $\mathcal{WQ}$ to $\mathcal{V}_1\mathcal{V}_2 \mathcal{V}_3$ for some additional quantum register $\mathcal{V}_3$.
If $\Psi$ is a hidden measurement channel for a projector $P$, then $\Psi'$ is obviously also a hidden measurement channel for $P$ for the partition $(\mathcal{V}_1 \mid \mathcal{V}_2 \mathcal{V}_3 )$.  The generalization follows.}.  For any $w \in \mathcal{W}$, let 
\begin{eqnarray}
r_w & = & U(w \otimes e_1) \\
s_w & = & U(w \otimes e_2) \\
R_w & = & \Mat ( r_w ) \\
S_w & = & \Mat ( s_w )
\end{eqnarray}
where $e_1,e_2$ are the standard basis vectors for $\mathcal{Q}$. Note that since $U$ is an isometry, $\left< R_w, S_{w'} \right> = 0$ for any $w,w' \in \mathcal{W}$. If
\begin{eqnarray}
P & = & \left[ \begin{array}{cc}
p & q \\ \overline{q} & t 
\end{array} \right]  \in \mathbb{S},
\end{eqnarray}
then $\Psi(ww^* \otimes P)$ can be written as
\begin{eqnarray*}
U\left(ww^* \otimes (pe_1e_1^* + q e_1 e_2^* + \overline{q} e_2e_1^* + t e_2 e_2^*) \right) U^* \\
= p r_w r_w^* + q r_w s_w^* + \overline{q} s_w r_w^* + t s_w s_w^*.
\end{eqnarray*}
Let $\rho(\cdot)$ be the marginal state of $\Psi( w w^* \otimes \cdot )$ on $\mathcal{V}_1$. Then, using Eq.~(\ref{eq: partial trace of second system}), we have that
\begin{eqnarray}
\rho(P) &= p R_w R_w^* + q R_w S_w^*
+ \overline{q} S_w R_w^*
+ t S_w S_w^*,  \hspace{0.66cm} \label{first marginal of P} \\
\rho(\mathbb{I}-P) &= t R_w R_w^* - q R_w S_w^*
- \overline{q} S_w R_w^*
+ p S_w S_w^*,  \hspace{0.66cm} \label{first marginal of I-P}
\end{eqnarray}
where for Eq.~(\ref{first marginal of I-P}) we also used the fact that $P$ has trace $1$ to simplify the coefficients.
Similarly if we let $\sigma(\cdot)$ be the marginal state of $\Psi(w w* \otimes \cdot )$ on $\mathcal{V}_2$, then by using Eq.~(\ref{eq: partial trace of first system}) we get that
\begin{eqnarray}
\sigma(P) = p \overline{R_w^* R_w} + q \overline{R_w^* S_w}
+ \overline{q} \overline{S_w^* R_w}
+ t \overline{S_w^* S_w}, \hspace{0.66cm} \label{second marginal of P} \\
\sigma(\mathbb{I}-P)= t \overline{R_w^* R_w} - q \overline{R_w^* S_w}
- \overline{q} \overline{S_w^* R_w}
+ p \overline{S_w^* S_w}. \hspace{0.66cm} \label{second marginal of I-P}
\end{eqnarray}
If $P$ is the orthogonal projector onto a unit vector $(x,y) \in \mathbb{C}^2$, then it can be written as
\begin{eqnarray}
P & = & \left[ \begin{array}{cc}
x \overline{x} & x\overline{y} \\ \overline{x}y & y\overline{y} 
\end{array} \right]
\end{eqnarray}
and hence equations~(\ref{first marginal of P})-(\ref{second marginal of I-P}) can be simplified to
\begin{eqnarray}
    \rho(P) &= (xR_w +yS_w)(xR_w+ yS_w)^* \label{eq: rho(P) for (x,y)} \\
    \rho(I - P) &= (\overline{y} R_w - \overline{x} S_w )(\overline{y} R_w - \overline{x}S_w)^* \label{eq: rho(I-P) for (x,y)} \\
    \sigma(P) &= (\overline{x} \overline{R_w} + \overline{y} \overline{S_w})^* (\overline{x} \overline{R_w} + \overline{y} \overline{S_w}) \label{eq: sigma(P) for (x,y)} \\
    \sigma(\mathbb{I} - P) &= (y \overline{R_w} - x \overline{S_w} )^* (y \overline{R_w} - x\overline{S_w}). \label{eq: sigma(I-P) for (x,y)}
\end{eqnarray}

Since $\Psi ( w w^* \otimes \left( \cdot \right) )$ is a hidden measurement channel for $\{ P , \mathbb{I} - P \}$ then the supports of $\rho(P)$ and $\rho(I-P)$ are orthogonal to each other, and the supports of $\sigma(P)$ and $\sigma(I-P)$ are orthogonal.
Thus we get the following equations:
\begin{eqnarray}
\label{hiddeneq1}
(xR_w + yS_w)(-\overline{y}R_w + \overline{x}S_w)^* & = & 0, \label{eq: hidden channel crit for RS*} \\
\label{hiddeneq2}
(x R_w + y S_w)^*(-\overline{y}R_w + \overline{x} S_w) & = & 0 \label{eq: hidden channel criteria for R*S}.
\end{eqnarray}
We now provide an alternative criteria for the hidden measurement channel condition --- one which is expressed directly in terms of the entries of $P$.
First notice that equations (\ref{eq: hidden channel crit for RS*}) and (\ref{eq: hidden channel criteria for R*S}) can be rewritten as
\begin{eqnarray*}
    \left[ \begin{array} {cc}
x  \mathbb{I} & y \mathbb{I}
    \end{array}
    \right]
    \left[ \begin{array} {cc}
R_w R_w^* & R_w S_w^*\\
S_w R_w^* 
 & S_w S_w^*
    \end{array}
    \right]
    \left[ \begin{array} {c}
-y \mathbb{I} \\
x \mathbb{I}
    \end{array}
    \right] & = & 0,
\end{eqnarray*}
\begin{eqnarray*}
    \left[ \begin{array} {cc}
\overline{x}  \mathbb{I} & \overline{y} \mathbb{I}
    \end{array}
    \right]
    \left[ \begin{array} {cc}
R_w^* R_w & R_w^* S_w\\
S_w^* R_w 
 & S_w^* S_w
    \end{array}
    \right]
    \left[ \begin{array} {c}
-\overline{y} \mathbb{I} \\
\overline{x} \mathbb{I}
    \end{array}
    \right] & = & 0.
\end{eqnarray*}
Then, by left multiplying the first equation (resp. the second equation) by $\left[ \begin{array} {cc}
\overline{x}  \mathbb{I} & \overline{y} \mathbb{I}
    \end{array}
    \right]^\top$ (resp. $\left[ \begin{array} {cc}
x  \mathbb{I} & y \mathbb{I}
    \end{array}
    \right]^\top$) and right multiplying it by $ \left[ \begin{array} {cc}
-\overline{y}  \mathbb{I} & \overline{x} \mathbb{I}
    \end{array}
    \right]$ (resp. $\left[ \begin{array} {cc}
x  \mathbb{I} & y \mathbb{I}
    \end{array}
    \right]$) we get

\begin{eqnarray*}
    \left[ \begin{array} {cc}
p  \mathbb{I} & \overline{q} \mathbb{I}\\
q \mathbb{I}
 & t   \mathbb{I}
    \end{array}
    \right]
    \left[ \begin{array} {cc}
R_w R_w^* & R_w S_w^*\\
S_w R_w^* 
 & S_w S_w^*
    \end{array}
    \right]
    \left[ \begin{array} {cc}
(1 - p) \mathbb{I} & - q \mathbb{I}\\
- \overline{q} \mathbb{I}
 & ( 1 - t ) \mathbb{I}
    \end{array}
    \right] & = & 0,
\end{eqnarray*}
\begin{eqnarray*}
    \left[ \begin{array} {cc}
p  \mathbb{I} & q \mathbb{I}\\
\overline{q} \mathbb{I}
 & t   \mathbb{I}
    \end{array}
    \right]
    \left[ \begin{array} {cc}
R_w^* R_w & R_w^* S_w\\
S_w^* R_w 
 & S_w^* S_w
    \end{array}
    \right]
    \left[ \begin{array} {cc}
(1 - p) \mathbb{I} & - \overline{q} \mathbb{I}\\
- q \mathbb{I}
 & ( 1 - t ) \mathbb{I}
    \end{array}
    \right] & = & 0.
\end{eqnarray*}

The proof of Theorem~\ref{thm:finitevalues} is based on the following two lemmas.

\begin{lemma}
\label{innerlemma}
Let $(P,v)$ and $(L,w)$ be elements of $\Lambda$.  Then,
\begin{eqnarray*}
\left\| v - w \right\| & \geq & \Omega \left( \sqrt{ \left\| P - L \right\|_1} \right).
\end{eqnarray*}
\end{lemma}

\textit{Proof.} By an appropriate change of basis on $\mathcal{Q}$, we assume without loss of generality that $P$ is the projector onto the vector $(1,0) \in \mathcal{Q}$ and $L$ is the projector onto the vector $(\cos \theta, - \sin \theta ) \in \mathcal{Q}$ for some $\theta \in [0, \pi/2]$.  
Let
\begin{eqnarray*}
     V  & = & R_v \\
    V' & = & S_v \\
    W & = & (\cos \theta) R_w - (\sin \theta) S_w \\
    W' & = & (\sin \theta ) R_w + (\cos \theta) S_w .
\end{eqnarray*}
The assumptions made so far imply the following conditions:
\begin{eqnarray*}
\left\| V \right\|_2 =
\left\| V' \right\|_2 =
\left\| W \right\|_2 =
\left\| W' \right\|_2 = 1 \\
\begin{array}{rl}
V (V')^* = 0,  & V^* ( V') = 0 \\
W (W')^* = 0, & W^* ( W') = 0
\end{array} \\
\left< V, W \right> =  (\cos \theta) \left< v, w \right> \\
\left< V', W' \right>  =  (\cos \theta) \left< v, w \right> \\
\left< V, W' \right>  =  (\sin \theta) \left< v, w \right> 
\end{eqnarray*}
Let $B \colon \mathcal{W} \to \mathcal{W}$ denote orthogonal projection onto the support of $V V^*$ (which is orthogonal to the support of $V'(V')^*$), and let $B^\perp = \mathbb{I} - B$.  
We have the following, in which we apply the equalities $B^\perp V = 0$, $W (W')^* = 0$, $(V')^* B = 0$, and the Cauchy-Schwarz inequality:
\begin{eqnarray*}
   \Re  \left< V, W' \right>  & = &
   \Re  \ls \Tr [ V (W')^* ]  \\
& = &   \Re \ls \Tr [ B V (W')^* B +
    B^\perp V (W')^* B^\perp ]  \\
& = &   \Re \ls  \Tr [ B V (W')^* B  ]  \\
& = &   \Re \ls \Tr [ B (V-W) (W')^* B  ] \\
& = &  \Re \ls  \Tr [ B (V-W) ( W' - V' )^* B ]   \\
& \leq & \left\| B ( V - W ) \right\|
\left\| (W' - V')^* B \right\| \\
& \leq & \left\|  V - W  \right\|
\left\| W' - V' \right\| \\
& = & \sqrt{ ( 2 - 2 \Re \left< V,W \right>  )
(2 - 2 \Re \left< V', W' \right> )}.
\end{eqnarray*}
Therefore,
\begin{eqnarray}
    (\sin \theta) \Re  \left< v, w \right>  & \leq &  2 - 2 (\cos \theta)  \Re \left< v, w \right>
\end{eqnarray}
which implies
\begin{eqnarray}
\Re \left< v, w \right> & \leq & \frac{2}{\sin \theta + 2 \cos \theta } 
\end{eqnarray}
and
\begin{eqnarray}
\left\| v - w \right\| & = & \sqrt{ 2 - 2 \Re  \left< v , w \right> } \\
& \geq  & \sqrt{ 2 - \frac{4}{\sin \theta + 2 \cos \theta }  } \\
& \geq & \Omega ( \sqrt{ \theta} ) \\
& \geq & \Omega ( \sqrt{ \left\| P - L \right\|_1} ),
\end{eqnarray} as desired. $\Box$

\begin{lemma}
\label{lem:diffpath}
Let $\pi \colon [0, 1 ] \to \Lambda$ be an infinitely differentiable path in $\Lambda$, expressed as $\pi ( a ) = (\pi_1 ( a), \pi_2 (a)) \in \mathbb{S} \times \mathcal{W}$.  Then, $\pi_1(0) = \pi_1 (1)$.
\end{lemma}

\textit{Proof.} Let $C = \max_{0 \leq a \leq 1} \left\| \pi_2' ( a ) \right\|$.  Then, $\left\| \pi_2 ( a ) - \pi_2 ( b ) \right\|$ is always less than or equal to $C | a - b |$ for all $a, b \in [0, 1 ]$. By Lemma~\ref{innerlemma},
\begin{eqnarray}
\label{Cineq}
    \left\| \pi_1 (a ) - \pi_1 ( b ) \right\| & \leq & O ( C^2 (a - b)^2)
\end{eqnarray}
for any $a, b \in [0, 1 ]$.  For any positive integer $N$, applying inequality~(\ref{Cineq}) inductively with $a$ drawn from the set $\{ j/N \mid j \in \{ 0, 1, \ldots, N \} \}$ yields the inequality
\begin{eqnarray}
    \left\| \pi_1 (0 ) - \pi_1 ( 1 ) \right\| & \leq & O ( C^2 / N ).
\end{eqnarray}
Since this inequality holds for any $N$, we have $\pi_1 ( 0 ) = \pi_1 ( 1 )$. $\Box$

Now we will finish the proof of Theorem~\ref{thm:finitevalues}.  We begin by describing the set $\Lambda$ in terms of a system of polynomial equations.  
Recall that $X,Y,Z$ denote the Pauli operators on $\mathbb{C}^2$.
For any unit vector $k = (k_1, k_2, k_3) \in \mathbb{R}^3$, let 
\begin{eqnarray}
P_k & = & (\mathbb{I} + k_1 X + k_2 Y + k_3 Z )/2.
\end{eqnarray}
Based on the equations immediately above
Lemma~\ref{innerlemma}, 
the set $\Lambda$ can be described as the set of all pairs
$(P_k, w )$
with $k \in \mathbb{R}^3$ and $w \in \mathcal{W}$, such that $\left\| k \right\|^2 = \left\| w \right\|^2 = 1, $ and
\begin{eqnarray}
\label{algdefining}
\nonumber
(P_k^\top \otimes \mathbb{I}_{\mathcal{V}_1}) 
\left[ \begin{array} {cc}
R_w R_w^* & R_w S_w^*\\
S_w R_w^* 
 & S_w S_w^*
    \end{array}
    \right]
((\mathbb{I}_\mathcal{Q} - P_k) \otimes \mathbb{I}_{\mathcal{V}_1}) 
& = & 0 \\
\nonumber
(P_k \otimes \mathbb{I}_{\mathcal{V}_2}) 
\left[ \begin{array} {cc}
R_w^* R_w & R_w^* S_w\\
S_w^* R_w 
 & S_w^* S_w
    \end{array}
    \right]
((\mathbb{I}_\mathcal{Q} - P_k^\top) \otimes \mathbb{I}_{\mathcal{V}_2}) 
& = & 0.
\end{eqnarray}
Since $\Lambda$ is a real algebraic set, it can be expressed as the disjoint union of a finite number of sets $T_i \subseteq \mathbb{R}^3 \times \mathbb{C}^n$ each of which is diffeomorphic to a hypercube $(0,1)^{d_i}$ with $d_i \geq 0$ (see Theorem 5.38 in \cite{basu2006algorithms}).  Lemma~\ref{lem:diffpath} implies that on each set $T_i$, the $\mathbb{S}$-value is constant.  Therefore, there are only a finite number of $\mathbb{S}$-values occurring in $\Lambda$.  We can decompose $\Lambda$ into a disjoint union of a finite number of closed subsets of the form
\begin{eqnarray}
    \Lambda_P & = & \left\{ (P, w ) \in \Lambda \mid w \in \mathcal{W} \right\}.
\end{eqnarray}

As we observed above, $\Lambda$ is defined by a fourth-degree real polynomial system on $2n + 3$ variables, hence the Milnor-Thom theorem (Theorem 2 in \cite{milnor1964betti}) implies that the set $\Lambda$ has at most $4 \cdot 7^{2n+2}$ connected components.  Therefore, $\Lambda_P$ is nonempty for at most $4 \cdot 7^{2n+2}$ values of $P$. $\Box$



Finally, we state our result in terms of cheating strategies for the single-qubit measurement protocol.

\begin{theorem}
\label{thm:main}
    Suppose that $(\mathcal{A, B, C, D}, \psi, \{U_P \}, V )$ is a perfect cheating strategy for a subset $T \subseteq \mathbb{S}$. Let
    \begin{eqnarray}
        m & = & \dim \mathcal{A} 
        \cdot \dim \mathcal{B} \cdot \dim \mathcal{C}.
    \end{eqnarray}    
    Then, $\left| T \right| \leq 4 \cdot 7^{2m+2}$.  
\end{theorem}

\textit{Proof.}  For each $P \in T$, let $\psi_P \in \mathcal{A} \otimes \mathcal{B} \otimes \mathcal{C}$ be the state defined by $\psi_P = U_P \psi$.  Then, the states in (\ref{perfdist}) can be expressed as $V ( \psi_P \otimes P ) \textnormal{ and }
V ( \psi_P \otimes ( \mathbb{I} - P ))$. Since the cheating strategy is perfect, $V ( \psi_P \otimes ( \cdot ))$ is a hidden measurement channel (with respect to the partition $(\mathcal{AD} \mid \mathcal{BC})$) for all $P \in T$.  By Theorem~\ref{thm:finitevalues}, this is possible only if $\left| T \right| \leq 4 \cdot 7^{2m+2}$.  $\Box$

\section{Discussion}
While Theorem~\ref{thm:main} shows in principle that the single-qubit measurement protocol is resistant to attacks, it only rules out perfect cheating strategies.  (The theorem shows that adversaries with finite-dimensional quantum resources cannot simulate the behavior of a perfect prover.)
Since the result is not error-tolerant,  a natural next step would be to explore security results that are more practically significant.

Previous work on QPV has explored security against adversaries with bounded quantum resources.  
Bluhm et al.~\cite{bluhm2022single} showed  that by encoding the basis choice for the single-qubit measurement protocol into two $n$-bit strings (sent separately by the two verifiers), one obtains an error-tolerant QPV protocol
which is secure against adversaries whose local memory size is bounded by a linear function of $n$.
While \cite{bluhm2022single} only used two possible basis choices, follow-up work of Escol{\`a}-Farr{\`a}s and Speelman~\cite{escola2023single} found better performance by using a larger set of basis measurements. Several other works have studied the protocol of~\cite{bluhm2022single} under different security settings and resource limitations~\cite{allerstorfer2023making,asadi2024linear,asadi2024rank,allerstorfer2024relating}. 
Further research could explore whether better performance (including improved error tolerance) could be achieved by integrating the techniques in the aforementioned papers with the algebraic geometry approach used in this work.

\begin{acknowledgments}
\section{Acknowledgments}
The authors thank Yi-Kai Liu, Dustin Moody, and Harry Tamvakis for their help with this paper.  The opinions expressed in this paper are solely those of the authors, and do not reflect any official opinions or endorsements by NIST.
\end{acknowledgments}

\bibliography{PRA}

\end{document}